\documentclass[namedreferences]{solarphysics}

\usepackage[figuresright]{rotating}
\usepackage[hyperref,natbib,optionalrh,solaromanenum]{spr-sola-addons}
\usepackage{graphicx}
\usepackage{txfonts}
\usepackage{amssymb}
\usepackage{color}
\usepackage{booktabs}
\usepackage{nicefrac}

\usepackage{caption}
\DeclareCaptionLabelSeparator{myspace}{\rule{4pt}{0pt}}
\hypersetup{
    final=true,
    pageanchor=true,
    colorlinks=true,
    breaklinks=true,
    linkcolor=blue,
    citecolor=blue,
    urlcolor=blue,
    pdfpagemode=UseNone,
    pdftitle={Background-subtracted Solar Activity Maps},
    pdfauthor={Denker and Verma},
    pdfsubject={solar physics},
    pdfkeywords={Active Regions; 
        Solar Cycle, Observations;
        Magnetic fields, Photosphere; 
        Chromosphere;
        Instrumentation and Data Management}}

\newcommand\arcsec{\ensuremath{^{\prime\prime}}}

\definecolor{myGreen}{cmyk}{0.992,0.,0.083, 0.525}

\definecolor{myDarkRed}{rgb}{0.698, 0.094, 0.133}

\definecolor{ed}{rgb}{0.,0.254,0.469}

\begin{document}

\begin{article}

%
%

\begin{opening}

%
%

\title{Background-subtracted Solar Activity Maps}

\author[addressref={aff}, corref, email={cdenker@aip.de}]
    {\inits{C.\ }\fnm{C.\ }\lnm{Denker}\orcid{0000-0002-7729-6415}}
\author[addressref={aff}]
    {\inits{M.\ }\fnm{M.\ }\lnm{Verma}\orcid{0000-0003-1054-766X}}

\address[id=aff]{Leibniz-Institut f{\"u}r Astrophysik Potsdam (AIP), 
    An der Sternwarte 16, 14482 Potsdam, Germany}

\runningauthor{C.\ Denker and M.\ Verma}
\runningtitle{Background-subtracted Solar Activity Maps}

%
%

\begin{abstract}
We introduce the concept of a \textit{Background-subtracted Solar Activity Map} 
(BaSAM) as a new quantitative tool to assess and visualize the temporal 
variation of the photospheric magnetic field and the UV $\lambda 160$~nm 
intensity. The method utilizes data of the \textit{Solar Dynamics Observatory} 
(SDO) and is applicable to both full-disk observations and regions-of-interest. 
We illustrate and discuss the potential of BaSAM resorting to datasets 
representing solar minimum and maximum conditions: (1) Contributions of 
quiet-Sun magnetic fields, \textit{i.e.} the network and (decaying) plage, to 
solar activity can be better determined when their variation is measured with 
respect to the background given by ``deep'' magnetograms. (2) Flaring and 
intermittent brightenings are easily appraised in BaSAMs of the UV intensity. 
(3) Both magnetic-field and intensity variations demonstrated that the flux 
system of sunspots is well connected to the surrounding supergranular cells. 
In addition, producing daily full-disk BaSAMs for the entire mission time of 
SDO provides a unique tool to analyze solar cycle variations, showing how 
vigorous or frail are the variations of magnetic-field and intensity features.
\end{abstract}
\end{opening}

%
%

\section{Introduction}\label{SEC1} 

The long-term variation of the solar magnetic field was extensively studied 
using synoptic maps created from full-disk magnetograms. \citet{Gaizauskas1983} 
utilized synoptic maps of photospheric magnetic fields using the Kitt Peak 
National Observatory's full-disk photospheric magnetograms for the ascending 
phase of solar cycle~21. In their synoptic maps, they followed the ``complexes 
of activity" \citep{Bumba1965} and noticed that these complexes of activity 
formed within a month, maintaining themselves by addition of fresh magnetic 
flux for 3\,--\,6 solar rotations. 
\citet{deToma2000} carried out a similar study but for the ascending phase of 
solar cycle~23. In addition to using magnetic synoptic charts, they used 
time-series of the 10.7~cm radio flux, sunspot numbers, and the Mg\,\textsc{ii}
chromospheric index for determining the origin of the two 
activity minima in 1996. The synoptic charts provided the details of the 
activity belt and activity nests, indicating preferred longitude bands 
where activity reoccurred. The properties of global magnetic evolution are, for 
example, needed to constrain flux transport dynamo models.

With the advent of digital imaging, time-series analysis became an important 
tool to obtain information about the variation, dynamics, and evolution of solar 
features. For example, extracting the intensity along a spatial slice at a given 
time from a time-sequence yields so-called space-time or time-slice diagrams. 
They are commonly used, \textit{e.g}. to infer information about exploding 
granules \citep{Title1986}, to detect oscillatory motions of bright points in 
continuum images \citep{Wang1995}, to compare magnetic flux measurements derived 
from near-infrared and visible spectropo\-lar\-im\-e\-tric observations 
\citep{Lin1999}, and to determine the divergence of the horizontal velocity 
field \citep{Shine2000}. More recently, \citet{Verma2016} followed the complete 
evolution of an active region using space-time diagrams based on synoptic 
line-of-sight (LOS) magnetograms. Various other time-series analysis methods are 
widely used and implemented fostering a better understanding the physical 
processes on the solar surface, \textit{e.g.} difference maps \citep[see][for 
solar active region loops]{Aschwanden1999}, sliding averages \citep[see][for 
umbral flashes and running penumbral waves]{RouppevanderVoort2003}, time-lag 
maps \citep[see][for coronal loops]{Viall2012}, decorrelation times for 
lifetimes of flows and of active region magnetic structures 
\citep{Welsch2012, Verma2012b}, and spatial correlation analysis 
\citep[see][for an adaptation of \textit{Local Correlation Tracking} 
(LCT)]{Verma2011}. Although many of these methods are versatile, not all of them 
are suitable for determining the global or large-scale evolution of solar 
intensity and magnetic field.

To visualize variations of the magnetic field in and around a decaying 
sunspot, \citet{Verma2012a} presented a map of temporal variations in the 
magnetic flux above and below the local background, \textit{i.e.} the long-term 
average of the magnetic field. Analyzing a 12-hour 
time-series of LOS magnetograms revealed regions of enhanced activity. A 
spoke-like structure was discovered in the background-subtracted variation map, 
which indicated moving magnetic features (MMFs) emanating from the photometric
sunspot border, traveling along preferential paths, and reaching all the way to
the surrounding supergranular boundary. \citet{Kummerow2015} and
\citet{Verma2018} extended this work by computing a large sample of these
background-subtracted variation maps for various sunspots complemented by
time-series of UV images. While these studies were focused on a specific
region-of-interest (ROI) covering individual sunspots, we will carry our 
initial work forward and propose in this study efficient tools to infer
properties of the global magnetic field and the solar activity in general.

\begin{figure}[t]
\centering
\includegraphics[width=\textwidth]{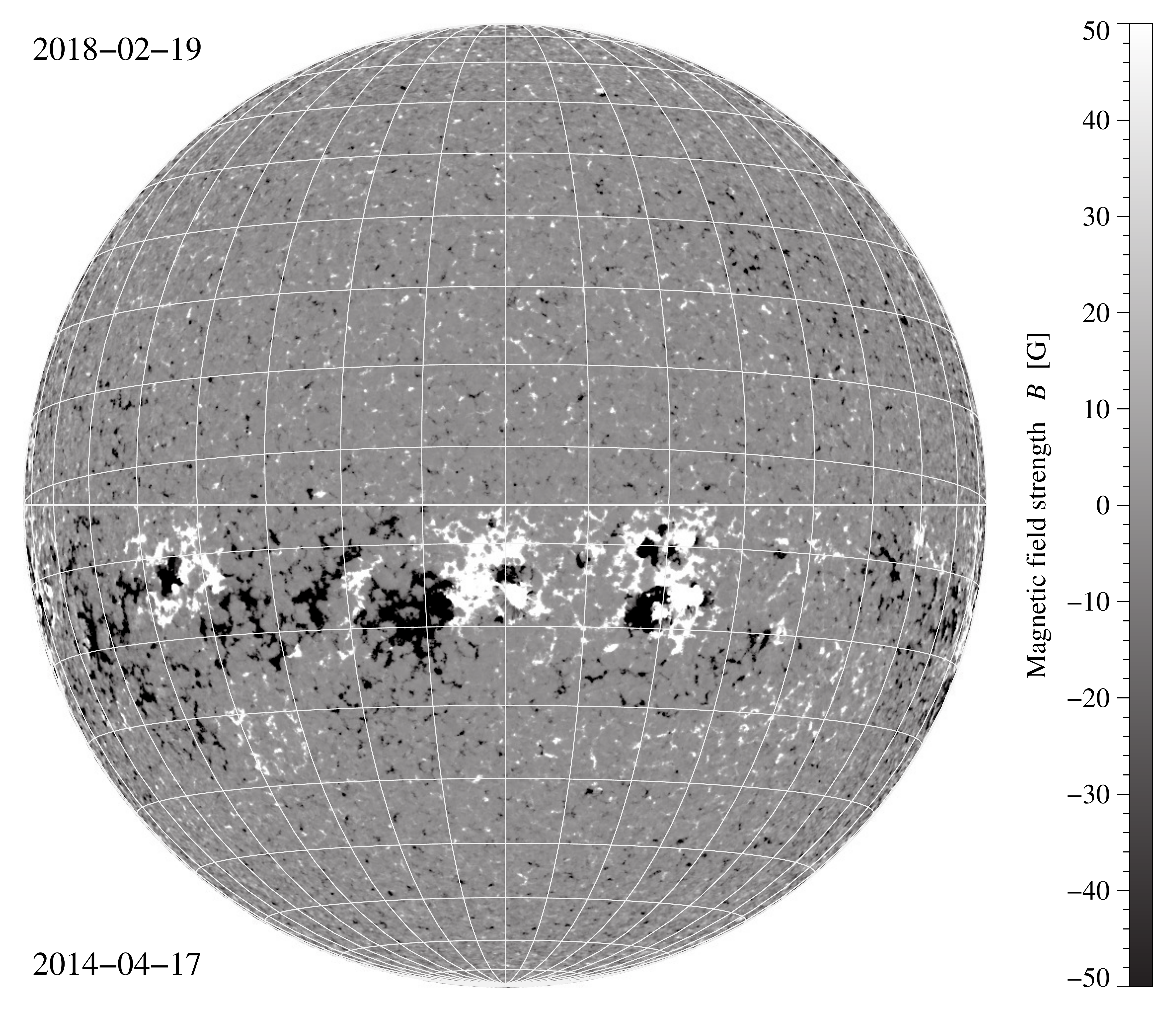}
\caption{Composite of two deep full-disk magnetograms $\langle B
    \rangle_\mathrm{16h}$ depicting solar minimum (\textit{top}) and maximum
    (\textit{bottom}) activity levels on 2018 February~19 and 2014 April~17,
    respectively.} 
\label{FIG01}
\end{figure}

In \citet{Beauregard2012}, we extended our implementation of LCT 
\citep{Verma2011}, originally developed by \citet{November1988}, to full-disk 
continuum images, and we used the \textit{Differential Affine Velocity 
Estimator} \citep[DAVE,][]{Schuck2005, Schuck2006} to derive flux transport
velocities from full-disk magnetograms. Large volumes of data are involved 
in computing horizontal flow fields. Having these three-dimensional data cubes
in hand (two spatial and the time coordinate) motivated us to explore the 
temporal variation of magnetic and UV activity for each pixel in the
field-of-view (FOV), complementing optical flow techniques. All methods 
mentioned above deal with images and magnetograms one way or the other. 
The purpose is to extract as much information as possible regarding solar 
activity and evolution of solar features from these kinds
of datasets. In the present study, we extend our previous work related to the
variation of magnetic fields and UV intensity \citep[\textit{e.g.}][]{Verma2012a,
Verma2018} and formally introduce the method as \textit{Background-subtracted
Solar Activity Map} (BaSAM).

In the following, we describe typical datasets (Section~\ref{SEC2}), present 
briefly the straightforward, though computationally extensive, implementation of
our method (Section~\ref{SEC3}), show results for full-disk and ROI data (Section~\ref{SEC4}), and discuss some of the common applications 
(Section~\ref{SEC5}).

\begin{figure}[t]
\centering
\includegraphics[width=\textwidth]{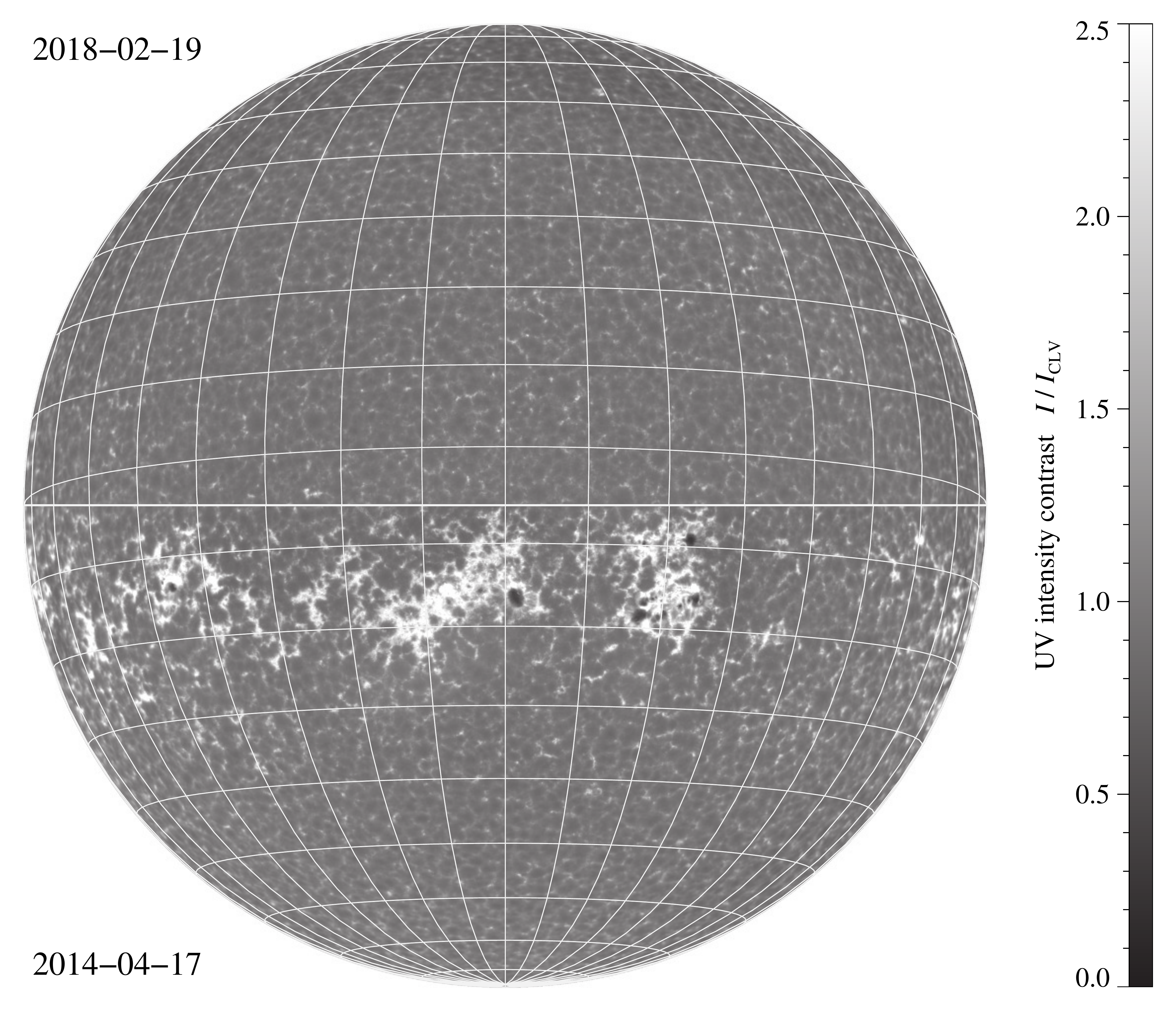}
\caption{Composite of two full-disk images of the limb-darkening
    corrected, normalized UV intensity $\langle I \rangle_\mathrm{16h}$.} 
\label{FIG02}
\end{figure}

%
%

\section{Observations}\label{SEC2} 

One day during solar maximum (2014 April~17) and another day during the 
declining phase (2018 February~19) of solar cycle 24 were selected for 
case studies to illustrate the potential of BaSAM (see Figures~\ref{FIG01} 
and \ref{FIG02}). In addition, a two-hour time window was selected for each 
day of the \textit{Solar Dynamics Observatory} \citep[SDO,][]{Pesnell2012}
mission so far (2010 May~1\,--\,2018 July~31), which resulted in a 
long-duration dataset containing about 3000 observing days. This 
long-duration dataset is the basis for the derivation of BaSAM indices 
tracing solar activity.

The activity maps are based on full-disk datasets of two SDO 
instruments, \textit{i.e.} the \textit{Helioseismic 
and Magnetic Imager} \citep[HMI,][]{Scherrer2012, Schou2012} and the
\textit{Atmospheric Imaging Assembly} \citep[AIA,][]{Lemen2012}. Time-series 
of line-of-sight (LOS) magnetograms and UV $\lambda 160$~nm images serve as
input for the case studies, which cover a period of $\pm 8$~hours around
12:00~UT. The cadence of HMI and AIA data are $\Delta t = 45$~s and 24~s,
respectively. Thus, 1280 magnetograms and 2400 UV images are potentially
available during the time period of $\Delta T = 16$~h. However, on 2018
February~19, only 1157 magnetograms were retrieved, \textit{i.e.} no 
magnetic field data were available for the time period from 06:42 to 08:08~UT.
In the UV time-series only about 10 images are missing, which is negligible.
All magnetograms and UV images in the time-series are compensated for
differential rotation using 12:00~UT as time reference. The long-duration
dataset comprises only magnetograms, which cover a period of $\pm 1$~hour
around 12:00~UT. However, eclipse seasons of SDO, when the view towards the 
Sun is obstructed, cause gaps in the time-series. These gaps were filled with
two-hour times-series of magnetograms that are as close in time to 12:00~UT 
as possible. While the bulk processing of the daily magnetogram data can be
automated, some manual interaction was needed to fill the gaps caused by the
eclipse seasons. 

The differential rotation is corrected using the standard
mapping routines of the SolarSoftWare library \citep[SSW,][]{Freeland1998}, 
which implements by default the differential rotation rate of small magnetic
features \citep{Howard1990b}. Compared to other differential rotation laws,
for example for different solar features or derived with other methods, 
the largest error is expected at the central meridian, where projection 
effects are minimal. For a rough error estimate, the default differential
rotation rate was compared to that given by \citet{Snodgrass1990}, which is 
based on photospheric Doppler velocity features. In the temporal sampling
window of one hour, the error is less than one third of a pixel, which is
negligible. Even for the longest sampling window of eight hours, the error
is only about 2.5~pixels. Hence, these mismatch errors may lead to some 
diffusion of the magnetic flux or intensity, which however can be safely
neglected.

Co-adding all magnetograms subsequent to differential-rotation correction
yields the deep magnetogram in Figure~\ref{FIG01}, where the magnetic field
strength $B$ was corrected for the cosine of the heliocentric angle $\mu =
\cos\theta$. This first-order correction ensures that magnetic fields near the
limb and disk center are equally well represented. However, in difference
images, this straightforward correction may lead to complications because
small differences will be strongly enhanced in proximity to the solar limb.
The deep full-disk magnetogram is a composite of the Sun's quiet northern
hemisphere on 2018 February~1 and the very active southern hemisphere on 2014
April~17. The northern hemisphere on 2014 April~17 is also very active, 
\textit{i.e.} it contains a large number of active regions, which are however 
not as prominent as those in the southern hemisphere. This type of composite
display allows us to directly compare activity levels at solar minimum and
maximum. The selected threshold of $\pm$50~G for the deep magnetogram enhances
both the network magnetic fields and the active plage regions. The superposed 
Stonyhurst grid represents solar longitude and latitude in $10^\circ$ 
increments for both observing days at 12:00~UT. The solid horizontal line 
does not mark the solar equator, it just separates the northern and southern
hemispheres. All full-disk data in the following sections adhere to the same
display style.

A long-integration UV intensity composite map was created using a similar 
procedure. However, we first determined the center-to-limb variation (CLV) on 
2018 February~19 by fitting a 4$^\mathrm{th}$-order polynomial in $\mu$ 
\citep[see][for a description of the procedure]{Denker1999a}, when the solar 
activity was very low, and divided the UV intensity map by a two-dimensional 
representation of the CLV. The same CLV correction was applied to the UV 
intensity map on 2014 April~17 after appropriate scaling, taking into account 
variations of the disk-center intensity. Thus, the composite of two full-disk 
images shown in Figure~\ref{FIG02} displays the UV intensity normalized with 
respect to the local quiet-Sun intensity. The imprint of the supergranulation, 
\textit{i.e.} the bright network (``orange peel pattern'') is much more 
pronounced in this normalized UV image as compared to the deep magnetograms. In 
any case, even though some structural contents is lost when taking the 
long-duration averages, some solar features will be enhanced so that this type 
of image processing has merits of its own. Apart from that further processing of 
time-series data comprised of magnetograms and UV images reveals additional 
information (see Section~\ref{SEC4}).

%
%

\section{Background-subtracted Solar Activity Maps}\label{SEC3}

\begin{figure}[t]
\centering
\includegraphics[width=\textwidth]{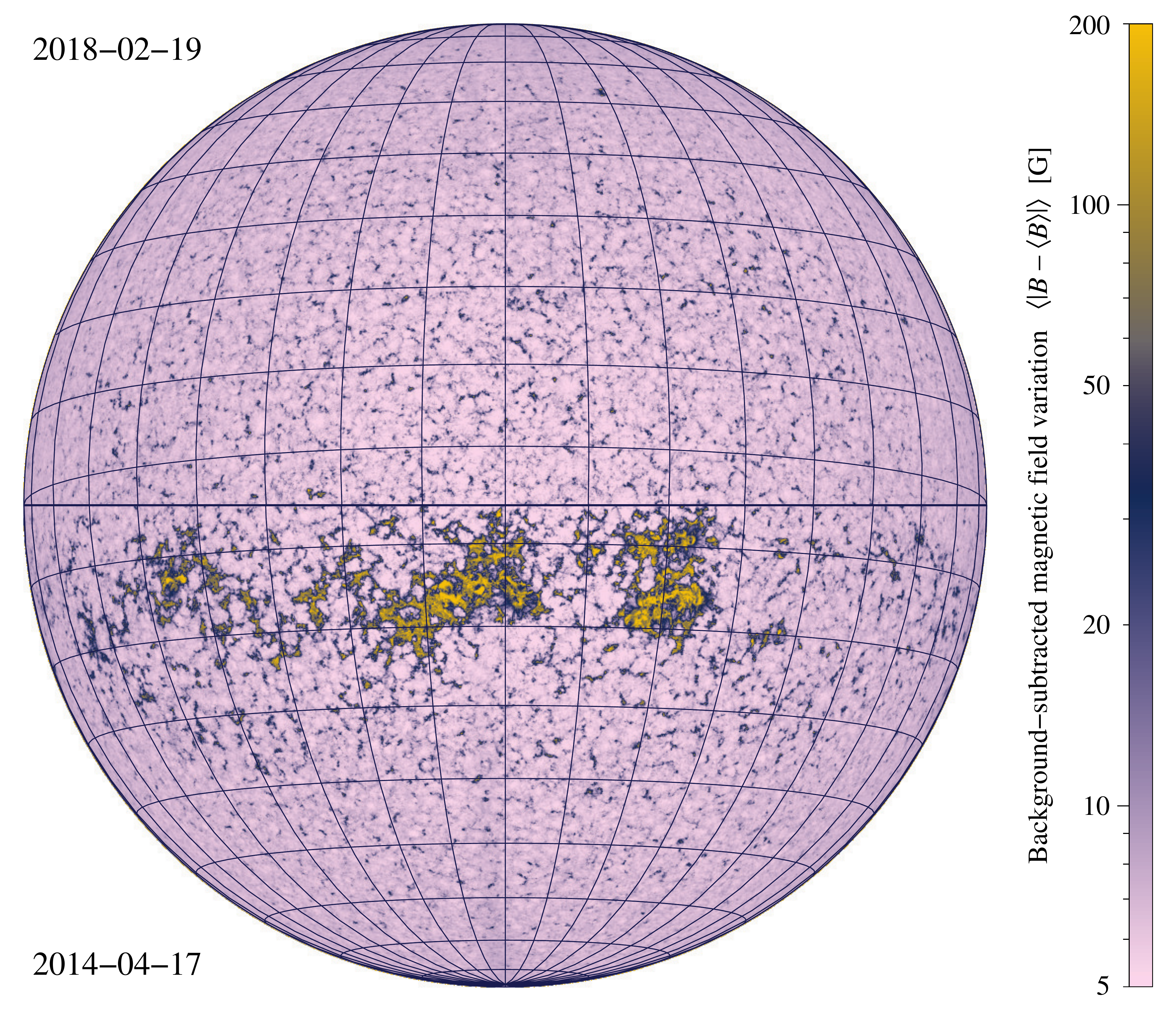}
\caption{Composite of two full-disk BaSAMs $\langle\,|\,B-\langle B\rangle\,|\,
    \rangle_\mathrm{16h}$ (logarithmic display) corresponding to
    Figure~\ref{FIG01}.} 
\label{FIG03}
\end{figure}

\begin{figure}[t]
\centering
\includegraphics[width=\textwidth]{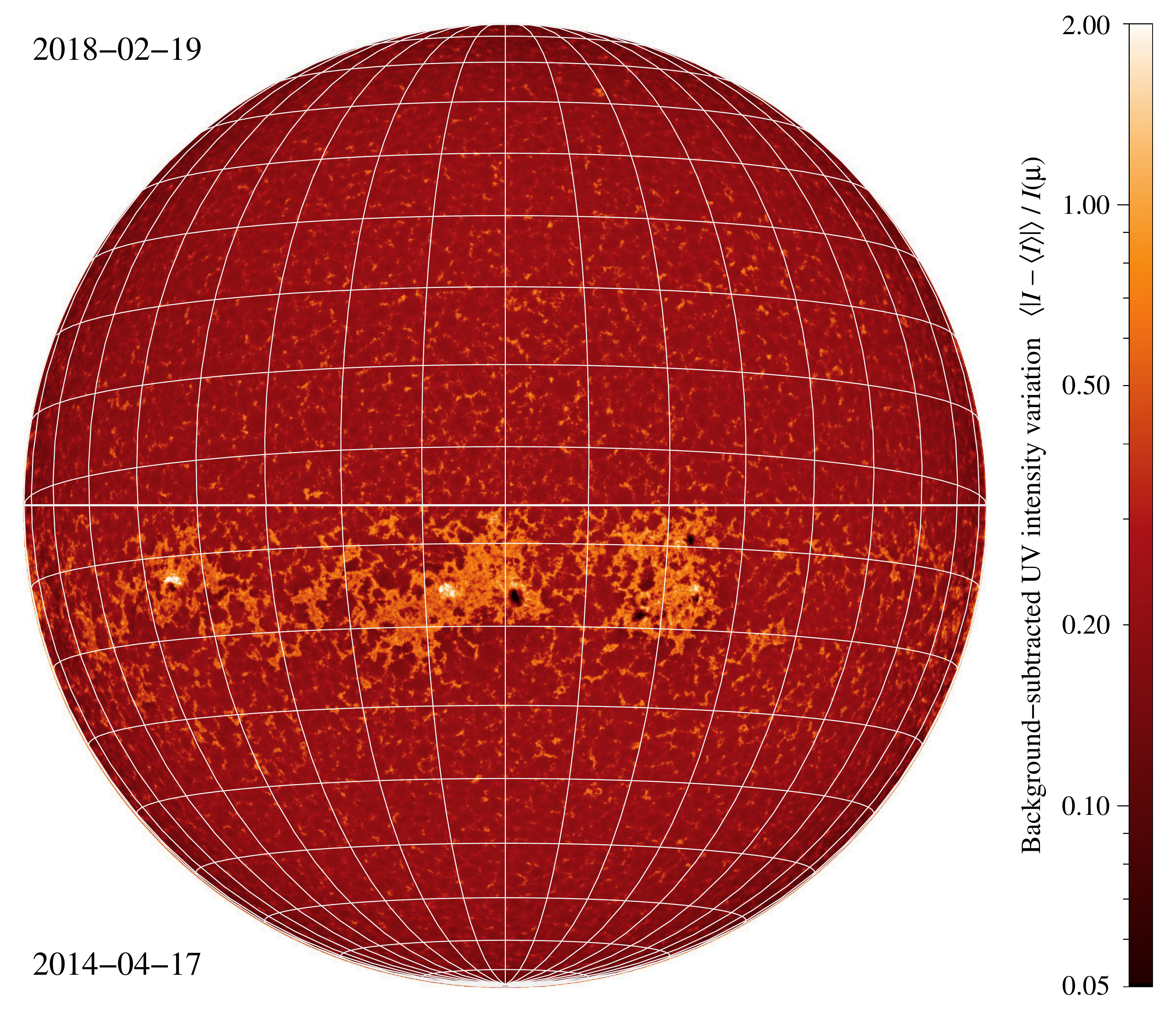}
\caption{Composite of two full-disk BaSAMs $\langle\,|\,I-\langle I\rangle\,|\,
    \rangle_\mathrm{16h} / I(\mu)$ (logarithmic display) corresponding to
    Figure~\ref{FIG02}.} 
\label{FIG04}
\end{figure}

The availability of high-cadence synoptic full-disk data with a moderate spatial 
resolution of about one second of arc is the prerequisite for our method to 
assess variations in magnetic field and UV/EUV imaging data. Instead of exploring 
the rms-contrast of two-dimensional surface data, we explore the temporal 
variation for each pixel on the solar disk. Thus, we implemented BaSAM as
the mean absolute deviation of a time-series calculated for each pixel on the
solar disk. In principle, the method can also be based on rms-measurements.
However, stronger variations will receive a higher weight in this case, which 
is undesirable when computing activity indices. In general, the
background-subtracted variation of a quantity $S$ with time, for example that
of the magnetic field strength $B$ or the UV intensity $I$, is computed 
according to
\begin{equation}
\langle\,|\,S-\langle S\rangle\,|\,\rangle = \frac{1}{N} \sum_{i=1}^{N} 
   \left|\,S(t_i) - \langle S\rangle\,\right| \quad \mathrm{with} \quad
   \langle S\rangle = \frac{1}{N} \sum_{i=1}^{N} S(t_i).
\label{EQN01}
\end{equation}
The individual images in a time-series are given by $S(t_i)$. The notation 
on the left-hand side of the equations is just shorthand for the mathematical
formalism on the right-hand side. The expression $\langle\,|\,B-\langle
B\rangle\,|\,\rangle_\mathrm{2h}$, for example, refers to a BaSAM based on 
a time-series with a duration of $\Delta T= 2$~h containing typically $N = 160$
magnetograms (Figure~\ref{FIG03}), whereas $\langle B\,\rangle_\mathrm{16h}$ 
is an abbreviation for a sixteen-hour deep magnetogram (Figure~\ref{FIG01}).
Since the cadence of the UV images is about two times higher, the UV
BaSAM $\langle\,|\,I-\langle I\rangle\,|\,\rangle_\mathrm{2h}$ is based on 
$N = 300$ images (Figure~\ref{FIG04}).

In Equation~\ref{EQN01}, the local background $\langle S\rangle$, which 
was computed just once, across the entire time-series, may be replaced by a 
simple sliding average
\begin{equation}
\langle S\rangle^M = \frac{1}{M+1} \sum_{j=i-M/2}^{j=i+M/2} S(t_j),
\label{EQN02}
\end{equation}
which has to be updated for each time step $t_i$ in order to compute 
the absolute difference $\left|\,S(t_i) - \langle S\rangle\,\right|$ in 
Equation~\ref{EQN01}. Thus, $\langle\,|\,I-\langle 
I\rangle\,|\,\rangle_\mathrm{16\,h}^\mathrm{30\,min}$ refers to a 16-hour BaSAM 
of the UV intensity that was computed with a 30-minute sliding average. Using 
sliding averages is advantageous, for example, when studying transient events 
such as flares. The notation in Equation~\ref{EQN02} does not address how to 
compute the sliding averages at the beginning and end of the time-series. We 
simply opted for computing the sliding average based on images contained within 
a time-series of given duration, \textit{i.e.} no images before or after the 
time-series were included in the average. Thus, the duration of the sliding 
average is shorter at the beginning and end of the times-series, which is however 
negligible as long as the duration of the sliding average is much shorter than 
the duration of the time-series. 

\begin{figure}[t]
\centering
\includegraphics[width=\textwidth]{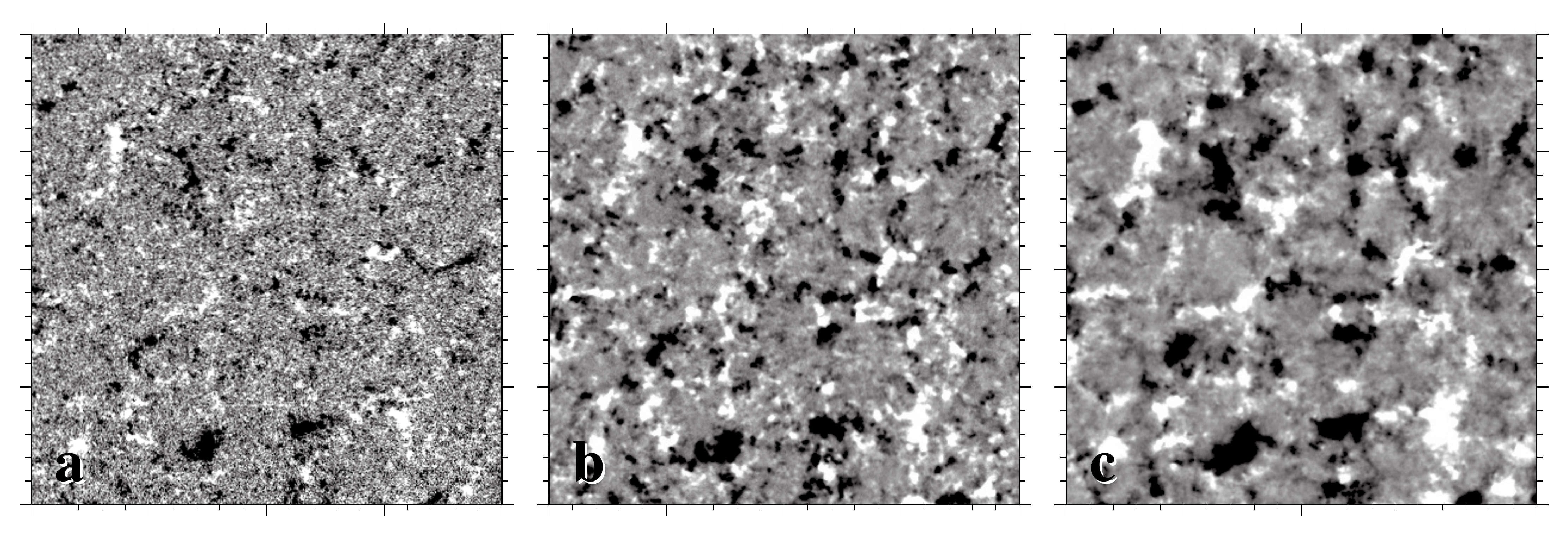}
\caption{Quiet-Sun magnetograms ($200\arcsec \times 200\arcsec$) taken at
    disk center at around 12:00~UT on 2018 February~19: (a) single, (b)
    2-hour average, and (3) 16-hour average magnetogram, respectively. The
    magnetograms were scaled such that 10\% of the pixels were clipped at 
    a threshold of $\pm$14.6~G, $\pm$8.5~G, and $\pm$6.4~G, respectively.} 
\label{FIG05}
\end{figure}

Photon noise is the dominant noise term for magnetograms and UV images.
Thus, the CLV introduces a systematic increase of the noise from disk center
to the limb. Table~1 in \citet{Couvidat2016} lists an uncertainty of 7~G for
the 45-second cadence, LOS magnetograms near disk center, which increases by
a factor of about 1.7 at the limb. Relevant information on the wavelength
dependent solar limb darkening is given by \citet{Pierce1977a} for HMI
magnetograms and by \citet{Pierce1977b} for AIA UV images, which can be used to
compute the noise level as a function of heliocentric angle. In addition,
solar differential rotation shifts the spectral line profile across the HMI
filter positions, which results in an additional noise component that increases
with distance from the central meridian. Since the noise in $\langle S \rangle$
is significantly lower than in $S(t_i)$, the latter will dominate 
$\left|\,S(t_i) - \langle S\rangle\,\right|$, which leads to a basal noise
floor in the summation of Equation~\ref{EQN01}. Moreover, magnetic BaSAMs and
deep magnetograms are governed by the same photon statistics based on the
number of ``integrated'' single magnetograms. The quiet-Sun magnetograms in
Figure~\ref{FIG05} visualize the noise being present in single and deep
magnetograms, respectively. The magnetograms were clipped at different 
thresholds ensuring that about 10\% of the pixels are saturated at the 
threshold. The single magnetogram only shows the strongest network elements, 
whereas the 2-hour averaged magnetogram shows both strong network and 
weak internetwork magnetic fields. The latter can lead to an additional 
basal component in magnetic BaSAMs based on very deep magnetograms (see
Figure~\ref{FIG05}c), where the contributions by weak internetwork fields
are washed out over the 16-hour averaging period.

In case studies, ROI processing is often more appropriate when focusing on objects 
such as pores, sunspots, or more complex active regions. However, additional
processing steps may be necessary: (1) Proper motions related to active region
evolution will be present in time-series. The cross-correlation between
single magnetograms and images and their respective long-duration averages
identifies these residual drifts, which are subsequently removed. (2) Resampling 
on a regular grid with an equidistant spacing will be advantageous if the 
same data are used to determine horizontal proper motions, which ensures that 
the sampling window to derive velocity vectors covers equal areas. (3) If the 
ROI is located in proximity to the solar limb, geometric corrections for the
magnetic field inclination or photometric corrections for the CLV will be
expedient. The data processing steps in these situations essentially follow 
the procedures elaborated in \citet{Verma2011} and \citet{Beauregard2012}. 
Thus, the choice of additional processing steps is driven by the specific 
science case. Only in statistical or comparative studies standardized 
processing steps are required.

%
%

\section{Results}\label{SEC4} 

Various parameters affect the appearance of BaSAMs. Thus, they must be 
carefully chosen according to the science case. In the following, we provide a
parameter study, investigating the impact of the duration for which BaSAMs are
computed and of the duration of the sliding averages. We selected an ROI in the
northern hemisphere (not shown in the composite full-disk BaSAMs in 
Figures~\ref{FIG01}\,--\,\ref{FIG04}) of the mature but decaying
$\beta\gamma$-region NOAA~12034, which showed minor flare activity in 
X-rays at the C-class level.

\begin{figure}[t]
\centering
\includegraphics[width=\textwidth]{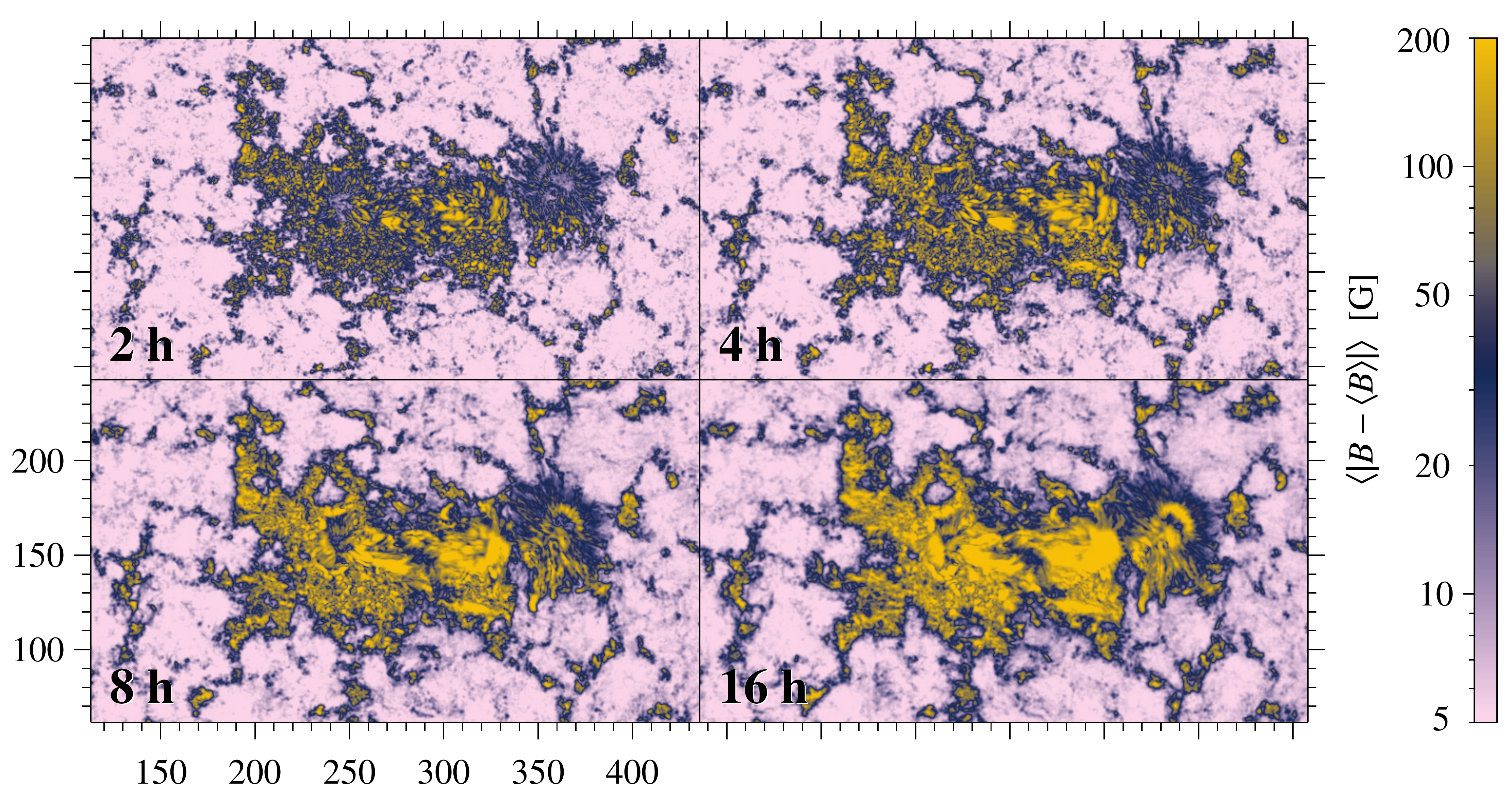}
\caption{Magnetic BaSAMs with a different duration of the background 
    $\langle\,|\,B-\langle B\rangle\,|\,\rangle_\mathrm{2h,\,4h,\,8h,\,16h}$
    (logarithmic display) showing active region NOAA~12034 on 2014 April~17. 
    The disc center coordinates are given in seconds of arc.} 
\label{FIG06}
\end{figure}

Varying the duration over which the background is computed affects
significantly the morphology of the BaSAM (see Figure~\ref{FIG06}). We
computed BaSAMs with a duration 2 to 16-hours. The significant morphological
changes that are apparent cannot be attributed to sparse temporal sampling.
Already for the 2-hour case, about 160 images are used to compute the BaSAM, 
which is sufficient to assume a Gaussian distribution for the magnetic field
variation in each pixel. The longer the duration, the smoother becomes 
the appearance of deep
magnetograms. Quiet-Sun magnetic fields become enhanced but with fuzzy
boundaries. As a consequence, most fine structure is visible in BaSAMs 
for backgrounds with a short duration because the background did not 
have time to evolve. This allows us to identify areas with pronounced
instantaneous magnetic field variations. In contrast, a longer duration 
for the background leads to larger variations of the magnetic field above 
a smoother background. Persistent variations stand out more clearly, 
which is advantageous when investigating magnetic connections within 
an active region or with its surroundings, \textit{e.g.} the magnetic 
network \citep[see][]{Verma2012a}.

\begin{figure}[t]
\centering
\includegraphics[width=\textwidth]{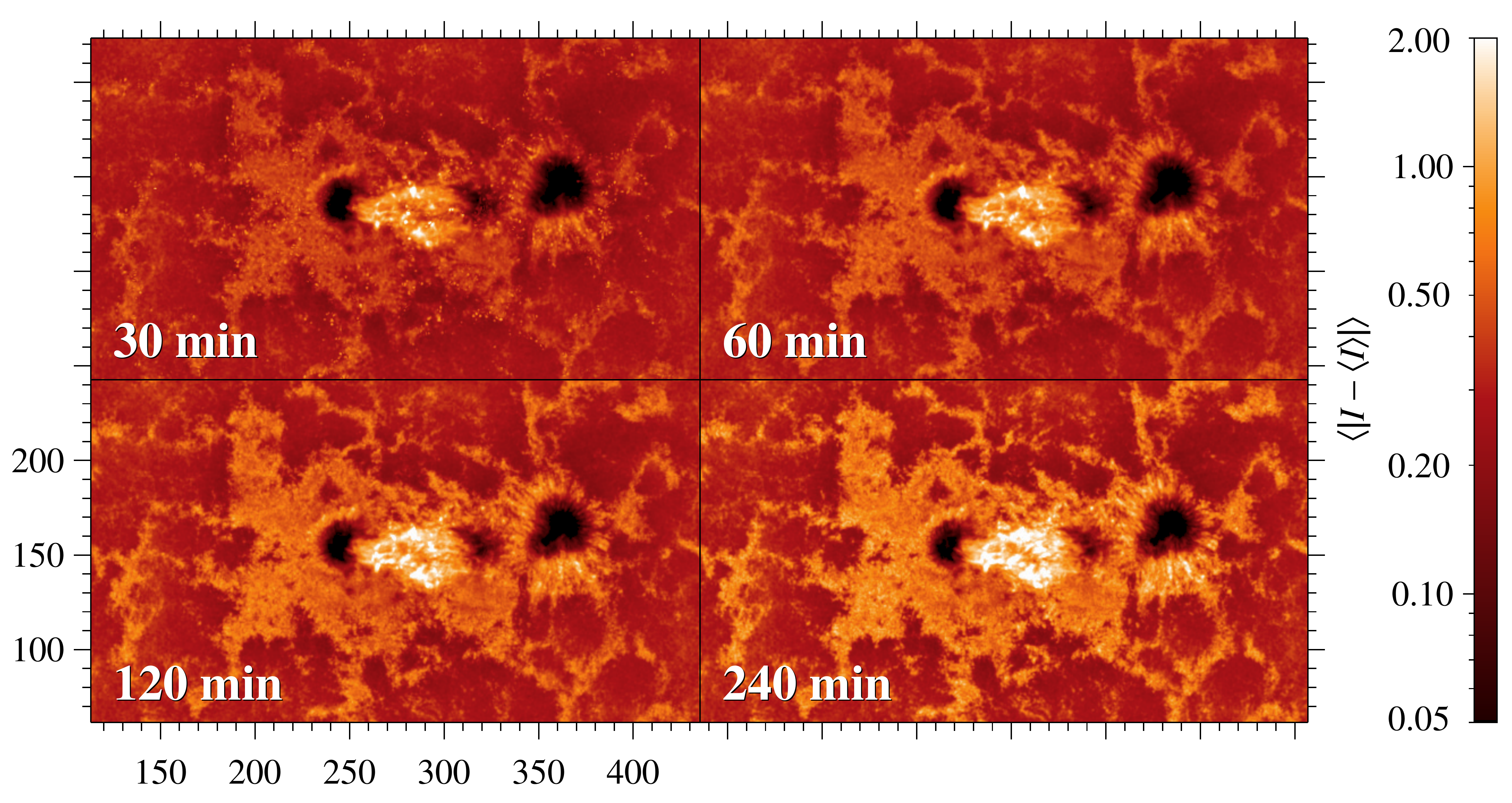}
\caption{UV BaSAMs with a different duration of the sliding average  
    $\langle\,|\,I-\langle
    I\rangle\,|\,\rangle_\mathrm{16h}^\mathrm{30min,\,60min,\,120min,\,240min}$
    (logarithmic display) for active region NOAA~12034 (see Figure~\ref{FIG06}).
    The duration over which the BaSAMs were computed was 16~hours in all cases.}
\label{FIG07}
\end{figure}

Using sliding averages for the background places emphasis on instantaneous
variations. The duration of the sliding averages impacts the appearance of
BaSAMs as shown for the UV intensity variation over a 16-hour period (see
Figure~\ref{FIG07}). The duration of the sliding average was doubled in each
step from 30~min, over 60 and 120~min, to 240~min. Even though no C-class 
flares or larger were recorded for active region NOAA~12034 on 2014 
April~17, all UV BaSAMs show clearly brightenings between opposite-polarity
sunspots at a location with mixed polarities. Interestingly, these 
brightening are absent around the leading sunspot. The strength of the
brightenings increases with the duration of the sliding average for the 
same reasons mentioned above for the magnetic BaSAMs (Figure~\ref{FIG06}). 
For the shortest sliding averages the brightenings decompose into point-like
features, indicating that small-scale flaring occurs at very localized 
regions, which are smeared out for longer duration sliding averages.

\begin{table}[t]
\centering
\caption{Linear (\textit{above diagonal}) and rank-order (\textit{below 
    diagonal}) correlation coefficients for the four BaSAMs in
    Figures~\ref{FIG06} (\textit{left}) and \ref{FIG07} (\textit{right}),
    respectively.}
\smallskip
\noindent
\begin{minipage}{0.4\textwidth}
\begin{tabular}{rcccc}
     &  \textbf{2~h} &  \textbf{4~h} &  \textbf{8~h} & \textbf{16~h}\rule[-2mm]{0mm}{3mm}\\
\textbf{2~h} &      & 0.92 & 0.80 & 0.71\\
 \textbf{4~h} & 0.92 &      & 0.91 & 0.80\\
 \textbf{8~h} & 0.86 & 0.93 &      & 0.91\\
\textbf{16~h} & 0.79 & 0.86 & 0.93 &     \\

\end{tabular}
\end{minipage}%
\begin{minipage}{0.6\textwidth}
\hspace*{2mm}
\begin{tabular}{rcccc}
        &  \textbf{30~min} &  \textbf{60~min} & \textbf{120~min} & \textbf{240~min}\rule[-2mm]{0mm}{3mm}\\
 \textbf{30~min} &         &    0.94 &    0.90 &    0.86\\
 \textbf{60~min} &    0.94 &         &    0.98 &    0.94\\
\textbf{120~min} &    0.89 &    0.98 &         &    0.98\\
\textbf{240~min} &    0.85 &    0.94 &    0.98 &        \\

\end{tabular}
\end{minipage}
\begin{tabular}{c}
\rule{0.9\textwidth}{0mm}
\end{tabular}\vspace*{-2mm}
\label{TAB01}
\end{table}

To quantify the differences between BaSAMs, when varying the overall
duration or the duration of the sliding average for the background, 
the linear and rank-order correlation coefficients were computed and 
summarized in Table~\ref{TAB01}. Linear and rank-order correlation 
coefficients evaluate either linear or monotonic relationships between two
continuous variables, respectively. In general, the rank-order correlation 
is less restrictive, which is helpful if no \textit{a priori} knowledge is
available about the exact data model. However, the high values for both types 
of correlation coefficients indicates that a linear model represents the 
data well. Using the 2-hour magnetic BaSAM as reference, the correlation
decreases with longer duration of the background. This decorrelation is
stronger for the rank-order correlation coefficient. In addition, 
consecutive maps show about the same values for linear as well as
rank-order correlation coefficients. The linear and rank-order 
correlation coefficients are about the same for the UV BaSAMs
with a different duration of the sliding averages. Furthermore, the
trend for the temporal decorrelation is similar to that of the magnetic
BaSAMs. In summary, the overall high correlation coefficients indicate 
that all BaSAMs capture the general properties of temporal variations 
in active regions. Furthermore, the duration of the background and the
duration of the sliding averages affect the fine structures contained
in the BaSAMs, \textit{i.e.} the parameters have to be adapted to the
scientific objective and the observed feature on the Sun.

\begin{figure}[t]
\centering
\includegraphics[width=\textwidth]{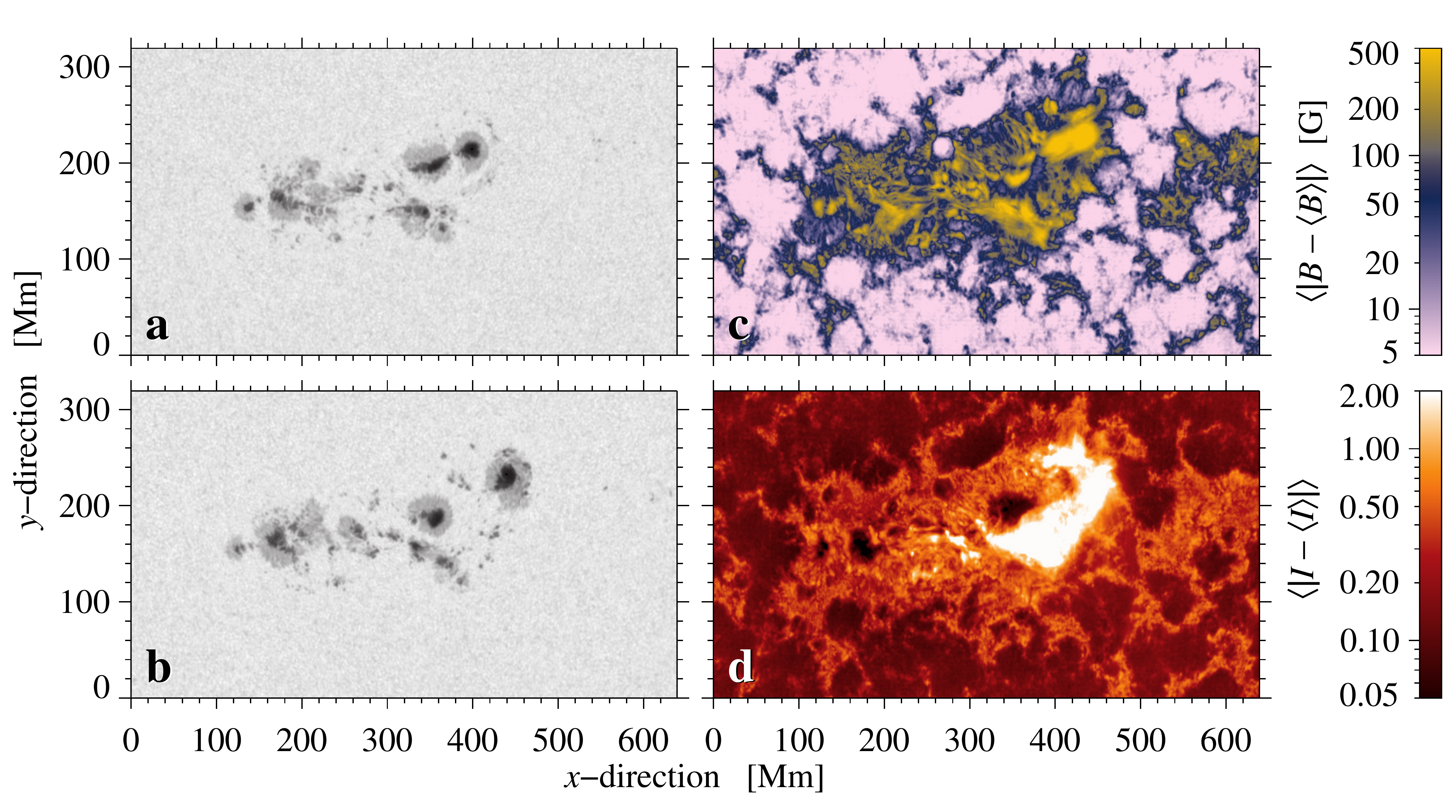}
\caption{BaSAMs of active region NOAA~11515 for (c) magnetograms 
    $\langle\,|\,B-\langle B\rangle\,|\,\rangle_\mathrm{24h}$ and 
    (d) UV images $\langle\,|\,I-\langle I\rangle\,|\,\rangle_\mathrm{24h}$.
    The fast evolution of the region is evident in the continuum images observed
    at 16:00~UT on (a) 2012 July~2 and (b) the following day, respectively.} 
\label{FIG08}
\end{figure}

We present two cases illustrating that BaSAMs are applicable to smaller ROIs: 
(1) the flare-prolific active region NOAA~11515 and (2) the active region NOAA~12081 
containing an axisymmetric sunspot. Differential rotation, CLV, and 
geometrical foreshortening were corrected for both ROIs 
\citep[see][]{Verma2018}. In NOAA~11515, we focus on the 24-hour time-period 
starting 16:00~UT on 2012 July~2 \citep{Louis2014}, when the leading spot 
exhibited strong rotation with respect to the other sunspots in the active 
region. Another distinguishing property was the fast separation of the 
leading and trailing sunspots. This complex active region was the source of many 
M-class solar flares. Rotation and separation can be traced in the 24-hour 
BaSAMs for magnetic field and UV intensity (Figure~\ref{FIG08}). The 
elongated region in the central part of the ROI shows an enhanced variation of 
the magnetic field and UV intensity, which highlights the stretching of the 
active region and the separating motion of the spots. This stretching is
more prominently visible in the magnetic BaSAM, where it appears as two 
elongated bands. The rotation of the leading sunspot becomes evident in the
UV BaSAM as a bright circular feature, which encompasses three quarters of 
the leading sunspot, in particular in penumbra locations. Axisymmetric 
sunspots may not produce clear signatures of rotation in BaSAMs. However, 
in the presence of asymmetric features such as a rudimentary/incomplete 
penumbra or the collision with encircling pores as in the case of the leading 
sunspot in active region NOAA~11515, rotation can be detected by the BaSAM
method. In particular, UV variations are caused by shear motions between 
leading sunspot and encircling pores, which led to continuous flickering
of the UV intensity, likely related to local reconnection.

\begin{figure}[t]
\centering
\includegraphics[width=\textwidth]{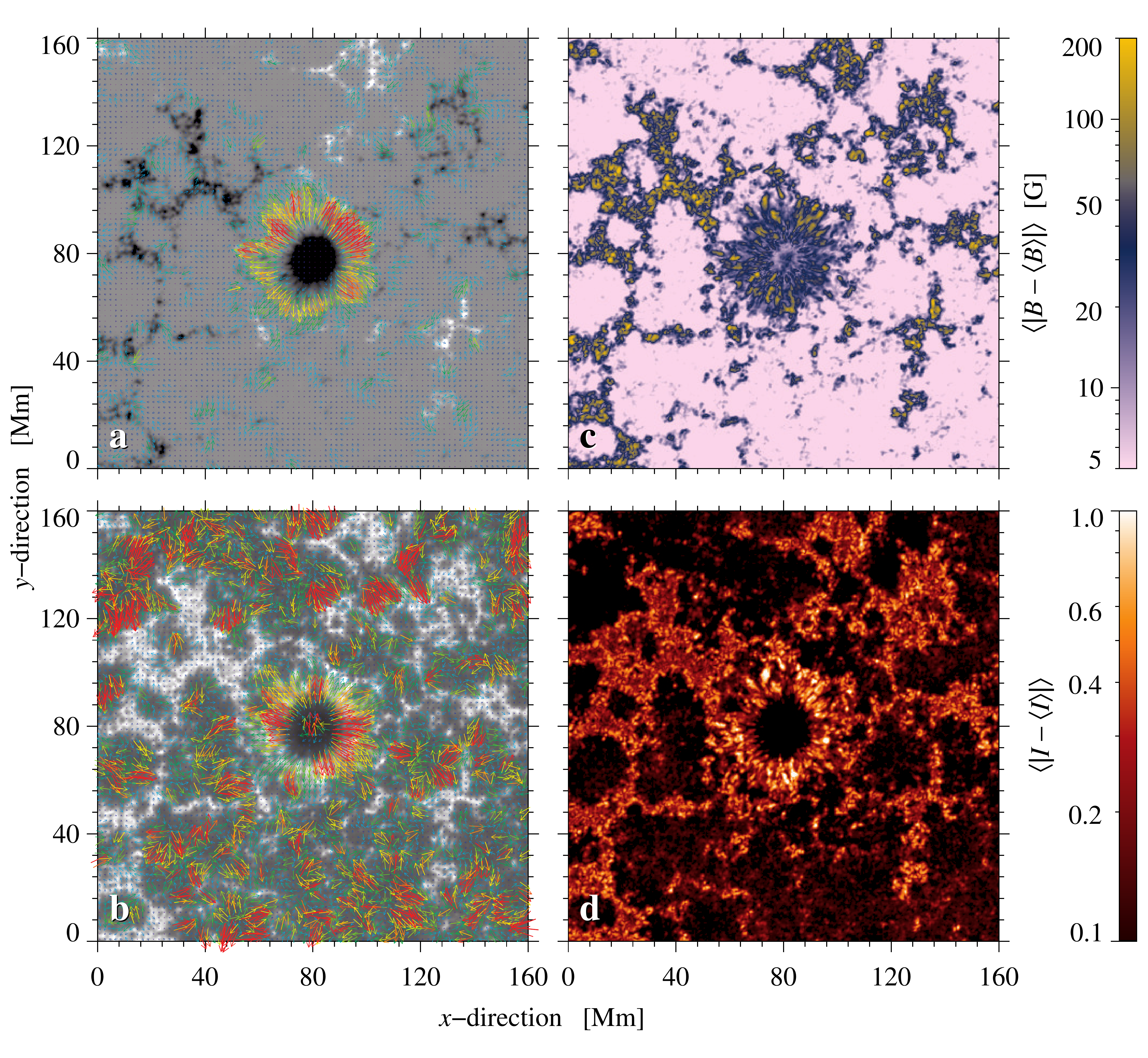}
\caption{Axisymmetric sunspot of active region NOAA~12081 with moat flows 
    measure using LCT and DAVE with corresponding BaSAM ROIs. The rainbow 
    colored arrows represent the magnitude as well as direction of the 
    horizontal proper motions. Violet and red arrows correspond to flow 
    speeds of lower than 0.1~km~s$^{-1}$ and larger than 0.4~km~s$^{-1}$,
    respectively.}
\label{FIG09}
\end{figure}

Apart from capturing splitting and shear motions in the more active regions, 
BaSAMs are a useful tool to investigate the connectivity of sunspots to their 
surroundings. Figure~\ref{FIG09} depicts flow maps of horizontal proper motions, 
which were computed using UV images and magnetograms, along with the 
corresponding BaSAMs for active region NOAA~12081 at the time close to its 
meridian crossing. The horizontal proper motions are estimated using LCT for 
the UV images and DAVE for the magnetograms. A detailed description of 
these two techniques that measure optical flows was presented in
\citet{Verma2018}. In the BaSAMs based on UV images and on magnetograms,
spoke-like structures emanating 
from the sunspot border reach the neighboring network. These structures coincide 
with regions exhibiting strong outward moat flows. Hence, it becomes apparent 
that the sunspot and the surrounding network are part of a larger magnetic flux 
system, which is conjoined by plasma flows. BaSAMs make this connection visible 
and help to investigate the nature of the interaction between magnetic fields 
and plasma motions.

\begin{figure}[t]
\centering
\includegraphics[width=\textwidth]{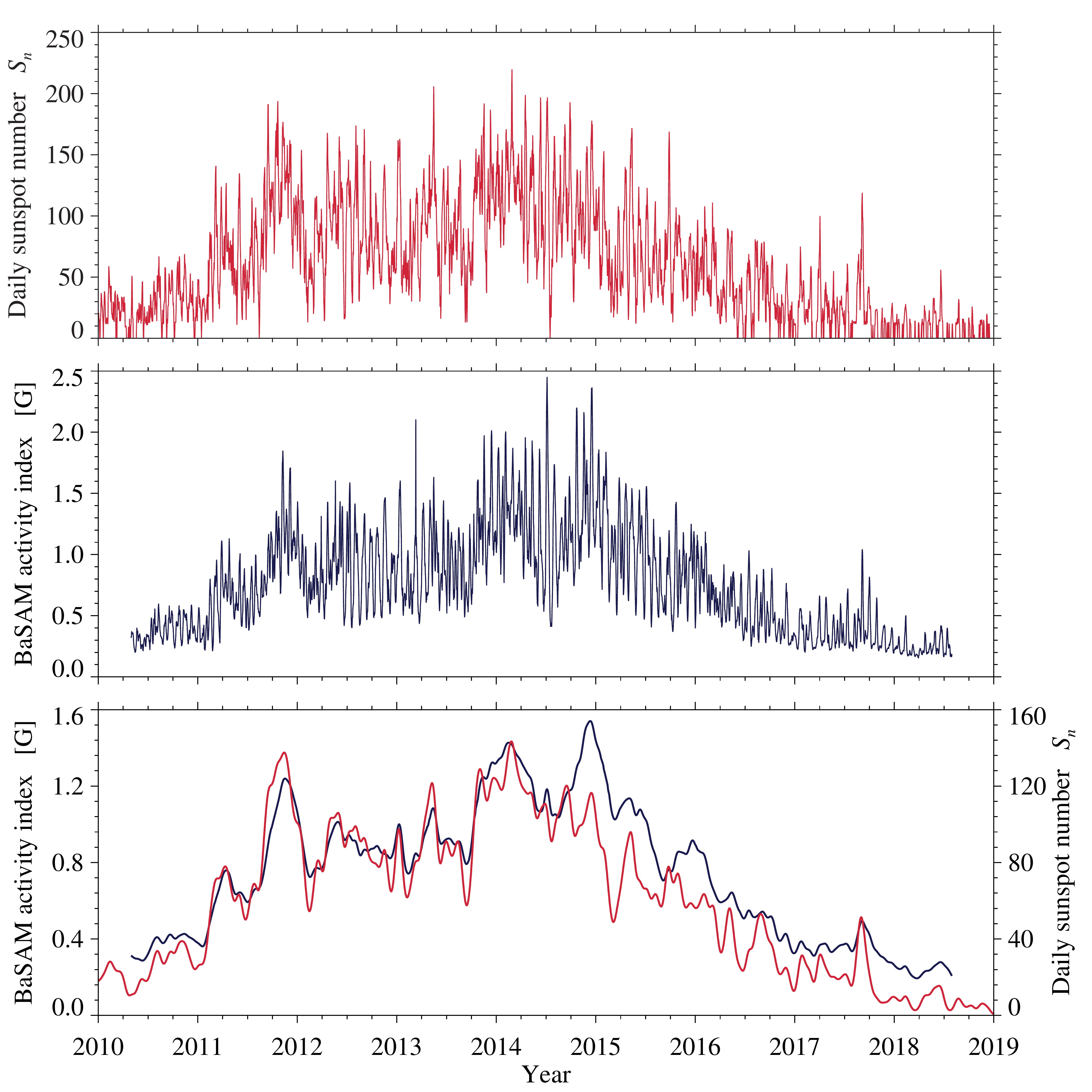}
\caption{Variation of the daily sunspot number $S_n$ (\textit{top}) during the
    9-year period since 2010. BaSAM activity index (\textit{middle}) using a
    threshold of 20~G for averaging the magnetic field variations across the
    solar disk. Small-scale variations in time reflect the synodic rotation 
    period of the Sun. In addition, the two maxima of solar cycle No.~24 are
    clearly visible in the BaSAM activity index. Smoothed versions 
    (Gaussian kernel with a FWHM = 28~days) of the sunspot number 
    (\textit{red}) and the BaSAM index (\textit{blue}) are displayed for
    direct comparison (\textit{bottom}).}
\label{FIG10}
\end{figure}

The availability of deep magnetograms and magnetic BaSAMs for every day 
since the start of the SDO mission opens the possibility to compute daily 
solar activity indices. The fact that about 160 magnetograms are condensed 
in a deep magnetogram or a magnetic BaSAM allows us to asses the contribution
of persistent small-scale magnetic fields to solar activity by simple
thresholding. The signal-to-noise ratio is increased in deep magnetograms
by a factor of more than twelve as compared to a single magnetogram,
where the noise level is about 7\,--\,12~G depending on the position on 
the solar disk. Using increasingly higher thresholds for deep magnetograms 
or BaSAMs changes the weight of quiet-Sun vs.\ active-region magnetic fields. 
In the middle panel of Figure~\ref{FIG10}, we show an activity index based on
magnetic BaSAMs, where we employed a threshold of 20~G for averaging the 
magnetic field variations across the solar disk, \textit{i.e.} the index 
provides the average variation of the magnetic field per pixel in a BaSAM. 
At the threshold level of 20~G, both quiet-Sun and active region magnetic
fields contribute to the magnetic field variations.
This approach is rather simplistic because it does not consider any geometric
correction with respect to the area contained in a single pixel or regarding 
the magnetic field inclination. However, already at this level solar
cycle properties and the solar rotation period become apparent, \textit{e.g.}
the double maximum of solar cycle No.~24.

The daily variation of the international sunspot number $S_n$ is plotted 
in the upper panel of Figure~\ref{FIG10} for comparison. While the sunspot 
number reaches zero in the absence of sunspots, \textit{i.e.} at the beginning
and end of the displayed time-series, the BaSAM index shows a distinct basal 
level for two reasons: (1) the presence of noise when computing the 
absolute mean deviation in Equation~\ref{EQN01} and (2) the existence of
ubiquitous magnetic fields on the solar surface -- even at
solar minimum -- and their variation. In addition, the BaSAM index appears 
smoother because it includes all magnetic fields that are present on the 
solar surface and not just those that leave a visible imprint in continuum 
or broad-band images such as sunspots and pores. Furthermore, the definition
of the sunspot number is based on the proper identification of individual 
sunspots and sunspot groups, which is prone to systematical errors, and 
results in integer values without any further subdivision. Consequently, 
the better statistical foundation of BaSAM activity indices makes them 
a good choice for time-series analysis.

The bottom panel of Figure~\ref{FIG10} shows the one-to-one comparison of the
BaSAM index and the sunspot number. Both curves are smoothed using a Gaussian
kernel with a FWHM of 28 days so that the two curves can be compared 
much more easily. These curves are very similar to the ones
presented in Figure~3 of \citet{Morgan2017} who discussed the global means
for coronal temperature, emission measure (EM), and photospheric magnetic
field for years 2010\,--\,2017. They found that active region temperature, EM, and
magnetic field are highly correlated. In addition, they find an initial peak in
early 2012 and a secondary peak in mid-2014 for some of these parameters before the
maximum in 2015. These features are also present in the BaSAM index but the
maximum in 2015 is missing in the daily sunspot number $S_n$. Interestingly,
the smoothed curves track each other very well up to this point, after which the 
BaSAM index displays a clear offset. The nature of this offset, whether
instrumental or real, still has to be investigated.

Time-series analysis using the Fourier transform of the activity index 
reveals a cluster of power peaks around the synodic
rotation period of the Sun. The periods of the three most significant peaks 
are in order of significance 26.1, 28.3, and 27.2~days, respectively. At 
this moment, the origin of this splitting is not clear. Since the input for
computing activity indices are full-disk magnetograms, computing the latitude
dependence of activity indices \citep[\textit{e.g.}][]{Knaack2005} 
or searching for active longitudes \citep[\textit{e.g.}][]{Balthasar2007a} 
becomes straightforward. A more detailed investigation, however, is beyond 
the scope of the current study and is deferred to a future publication. Such a 
study will also include solar activity indices based on BaSAMs of the 
UV intensity.

%
%

\section{Discussion and Conclusions}\label{SEC5} 

We present a method to visualize variations of the UV intensity and magnetic 
field. BaSAMs are derived not only for solar full-disk data but also for 
smaller ROIs. The observed temporal variations are visualized in the spatial 
domain. In addition, the disk-integrated variation can be derived and used 
as an indicator of long-term solar activity changes, which can be translated into 
UV or magnetic activity indices. Sunspots numbers, sunspot area, and the 10.7~cm radio 
flux are among the commonly used indices representing long-term solar activity. 
Solar cycle variations are often presented in terms of Zurich sunspot numbers 
\citep[\textit{e.g.}][]{Chernosky1958}. Furthermore, shape and strength of next 
solar cycle are often predicted in terms of sunspot numbers 
\citep{Hathaway1999}. Indices based on BaSAMs display a good correlation with 
the aforementioned solar activity indices and have an additional advantage, 
\textit{i.e.} they can be computed for features, where the strength of the 
variations significantly differs. Thus, contributions of quiet-Sun and highly 
active regions can be separated.

BaSAMs are complementary and provide additional scientific insight to existing
solar full-disk data. The shift-and-add method for short-exposure images is 
one of the early attempts to improve photometric accuracy while preserving
fine-structure when observing through a turbulent medium such as Earth's
atmosphere \citep{Hunt1983}. Thus, co-adding images or magnetograms to enhance
photometric accuracy or magnetic sensitivity is nothing new. For example,
magnetograms reaching polarization levels of $10^{-4}$ can be produced with 
fast polarimeters and cameras in combination with post-processing 
\citep[see][for an implementation at \textit{Big Bear Solar Observatory}
(BBSO)]{Wang1998b}. They found that 1000 and 100 integrations are needed
for quiet-Sun and active regions, respectively, to reach the accuracy of the
aforementioned polarization level. We follow the same principle when creating
deep magnetograms. Even though SDO data are not affected by atmospheric seeing, 
solar differential rotation must be nevertheless corrected before co-adding
magnetograms. In our deep magnetograms, we are able to detect very weak 
magnetic fields in the polar and internetwork regions. Sensitive and 
high-cadence magnetograms are a prerequisite for the BaSAM method. For example,
considering the image scale (2\arcsec~pixel$^{-1}$), cadence (96~min), 
and detector technology, creating BaSAMs based on synoptic full-disk 
magnetograms obtained with the \textit{Michelson Doppler Imager}
\citep[MDI,][]{Scherrer1995} on board the \textit{Solar and Heliospheric
Observatory} \citep[SoHO,][]{Domingo1995} was out of reach at that time.

The multiple wavelengths channels of AIA provide full-disk solar images 
covering various solar atmospheric layers. To extract more information from
these full-disk images, \citet{Viall2012} used a property of coronal loops, 
where the peak intensities in cooler atmospheric layers are reached at 
different times than those of hotter atmospheric layers. By measuring the time
lag of the intensity response in pairs of EUV channels for a duration of
24~hours, they demonstrated that the coronal plasma is very dynamic and 
evolves on different time scales. This time-lag method computes temporal
cross-correlations of intensity profiles on a pixel-by-pixel basis using
properly aligned time-series of AIA images taken in two channels. The results 
are maps identifying variations, periodic phenomena, and oscillations from 
lower to higher atmospheric layers. The BaSAM method differs from this approach
because it refers to just one atmospheric layer and visualizes the accumulated
strength of temporal variations. UV images and magnetograms do not just
represent different data types but they also sample two different atmospheric
layers, \textit{i.e.} upper photosphere/transition region and photosphere.
However, both methods serve as examples that full-disk SDO data can be used 
as a starting point for value-added data products such as UV and magnetic 
BaSAMs that complement each other. 

Time-series full-disk data corrected for solar differential rotation
facilitated LCT \citep[\textit{e.g.}][]{Beauregard2012, Verma2018} and were 
the point of departure for developing the BaSAM method. Recently,
\citet{Morgan2018} proposed a new method to follow continuous faint motions 
in the corona using AIA EUV images. The \textit{Time-Normalized-Optical-Flow}
(TNOF) method employs sliding averages and temporal
smoothing to remove the slowly varying background from a time-series of EUV
images for each pixel. The processed time-series data can then be analyzed 
with optical flow tracking techniques, which facilitates following coronal 
motions at sub-pixel resolution. However, in this case the magnitude of the
variation is not decisive but its temporal evolution, which yields the 
motion vector at each pixel. The high quality, cadence, and spatial 
resolution of SDO AIA and HMI data is the \textit{condicio sine qua non}
for methods such as TNOF and BaSAM.

However, stray light degrades SDO data and affects the photometric precision, 
\textit{i.e.} the readily available level~1 data are not corrected with an 
appropriate point spread function (PSF). Multiple approaches were suggested 
to estimate the stray light contamination and to correct HMI continuum and 
magnetograms \citep[\textit{e.g.}][]{Yeo2014, Couvidat2016, Criscuoli2017}.
Only recently the deconvolved continuum images became available for
download. Recently, \citet{DiazBaso2018} introduced the ``Enhance'' method, 
which uses two deep neural networks to produce deconvolved and super-resolved 
HMI continuum and magnetograms. The neural networks are trained on MHD 
simulations of active regions. Future versions of BaSAM could be based on 
enhanced or sharpened images and magnetograms, making it easier to 
discriminate between contributions from small- and large-scale features to
activity indices. 

So far BaSAMs are only computed for full-disk UV 
images and magnetograms besides some smaller ROIs for case studies. 
A possible extension of BaSAM is to apply it to 
data obtained at other wavelengths that cover chromosphere, transition 
region, and corona. The next 
step, is to compute BaSAMs for high-resolution images and magnetograms. One 
limitation of high-resolution data is temporal and spatial coverage. In this 
respect, data from space missions such as \textit{Hinode} \citep{Kosugi2007} 
and \textit{Interface Region Imaging Spectrograph} 
\citep[IRIS][]{DePontieu2014} are a good starting point because ground-based 
data are affected by seeing and the day-night cycle. Nevertheless,
image restoration in combination with 
adaptive optics occasionally produces hour-long time-series with 
diffraction-limited quality at meter-class solar telescopes. When computing 
BaSAMs, we assumed the background to be quasi-stationary, which becomes 
questionable for high-resolution photospheric and chromospheric observations. 
In this case, sliding averages are more appropriate for the computation of 
the background. Thus, increased spatial resolution will present us with 
additional information.  In summary, BaSAM is a powerful method to investigate 
the dynamic Sun in the space and time domains.

%
%

\begin{acks}
SDO HMI and AIA data are provided by the Joint Science Operations Center --
Science Data Processing. The international sunspot number is compiled by 
the \textit{Sunspot Index and Long-term Observations} (SILSO) project of the 
Royal Observatory of Belgium, Brussels. We thank Drs.\ Rainer Arlt and Vasyl
Yurchyshyn for their comments and suggestions, which helped to improve the
manuscript. This study was supported by grant DE~787/5-1 of the Deutsche
Forschungsgemeinschaft (DFG) and by the European Commission's Horizons 
2020 Program under grant agreement Number 824064 (ESCAPE -- European Science 
Cluster of Astronomy \& Particle physics ESFRI research infrastructures). 
\medskip

\noindent\textbf{Disclosure of Potential Conflicts of Interest}$\quad$ The
authors declare that they have no conflicts of interest.
\end{acks}

%
%


\end{article}

\end{document}